\documentclass[
 aip,
 amsmath,amssymb,
 reprint,%
 graphicx,%
]{revtex4-1}

\usepackage{graphicx}
\usepackage{dcolumn}
\usepackage{bm}

\usepackage[utf8]{inputenc}
\usepackage[T1]{fontenc}
\usepackage{mathptmx}
\usepackage{soul}
\usepackage{xcolor}

\begin{document}

\preprint{AIP/123-QED}

\title{Planar Laser Induced Fluorescence Mapping of a Carbon Laser Produced Plasma}

\author{R. S. Dorst}
\email{rdorst@physics.ucla.edu.}

\author{C. G. Constantin}
\affiliation{Department of Physics and Astronomy, University of California - Los Angeles, Los Angeles, California 90095, USA}

\author{D. B. Schaeffer}
\affiliation{Department of Physics and Astronomy, University of California - Los Angeles, Los Angeles, California 90095, USA}
\affiliation{Department of Astrophysical Sciences, Princeton University, Princeton, New Jersey 08540, USA}

\author{J. J. Pilgram}

\author{C. Niemann}
\email{cniemann@g.ucla.edu.}
\affiliation{Department of Physics and Astronomy, University of California - Los Angeles, Los Angeles, California 90095, USA
}%

\date{\today}

\begin{abstract}

We present measurements of ion velocity distribution profiles obtained by laser induced fluorescence (LIF) on an explosive laser produced plasma (LPP). The spatio-temporal evolution of the resulting carbon ion velocity distribution was mapped by scanning through the Doppler-shifted absorption wavelengths using a tunable, diode-pumped laser. The acquisition of this data was facilitated by the high repetition rate capability of the ablation laser (1 Hz) which allowed the accumulation of thousand of laser shots in short experimental times. By varying the intensity of the LIF beam, we were able to explore the effects of fluorescence power against laser irradiance in the context of evaluating the saturation versus the non-saturation regime. The small beam size of the LIF beam led to high spatial resolution of the measurement compared to other ion velocity distribution measurement techniques, while the fast-gated operation mode of the camera detector enabled the measurement of the relevant electron transitions.

\end{abstract}

\maketitle

\section{Introduction}
\label{sec:intro}

Laser induced fluorescence (LIF) is a non-perturbative optical technique that is capable of selectively measuring the ion velocity distribution function of a plasma\cite{Muraoka92,Goeck88} species with high spatial and temporal resolution. 

LIF has been shown previously to characterize the neutral densities in tokamak edge plasmas\cite{Gorb12}, measure temperatures of argon plasmas\cite{Bonde2018}, and measure oxygen concentration in fluid mixing experiments\cite{Goldsmith86}. Its application to a laser-produced plasma is a novel approach, turning it into a powerful tool to measure the evolution of various ion charge states of interest\cite{Bonda2012}. Determining the  velocity distribution of particles in a plasma is a critical - and challenging - step in understanding the complex laser-target interactions, plasma dynamics and plasma interactions with other systems, such as ambient gases, magnetic fields, or other plasmas.

LIF is distinct from other active optical plasma diagnostics, such as Thomson scattering and Raman scattering \cite{Kaloyan2021}, in that it can measure ion properties directly, instead of through measuring electron properties and inferring ion properties by invoking quasi-neutrality\cite{Hutchinson} in the case of non-collective optical Thomson scattering, or through \textit{a priori} knowledge of the distribution shape for collective optical Thomson scattering\cite{schaeffer_prl2019}. 

The ion velocity distribution function (VDF) is constructed incrementally by tuning the LIF probe beam to Doppler shifted absorption wavelengths over successive shots until the entire width of the velocity distribution has been scanned over and measured. This requires that the bandwidth of the LIF beam be much smaller than the width of the ion VDF in question.

In this paper we present the first measurements of ion dynamics in a carbon laser produced plasma (LPP) by means of a new application of a LIF technique in a high repetition rate (HRR) experiment. 

The paper is structured as follows: Section~\ref{sec:setup} details the experimental setup and diagnostics. We then describe the specifics of the LIF scheme in Sec.~\ref{sec:plif} as well as the procedures for analyzing the data. In Sec.~\ref{sec:results} we then present the experimental findings. Section~\ref{sec:summary} is a summary. 

\section{Experimental Setup and Design}
\label{sec:setup}

\begin{figure}
    \centering
    \includegraphics[scale=0.3]{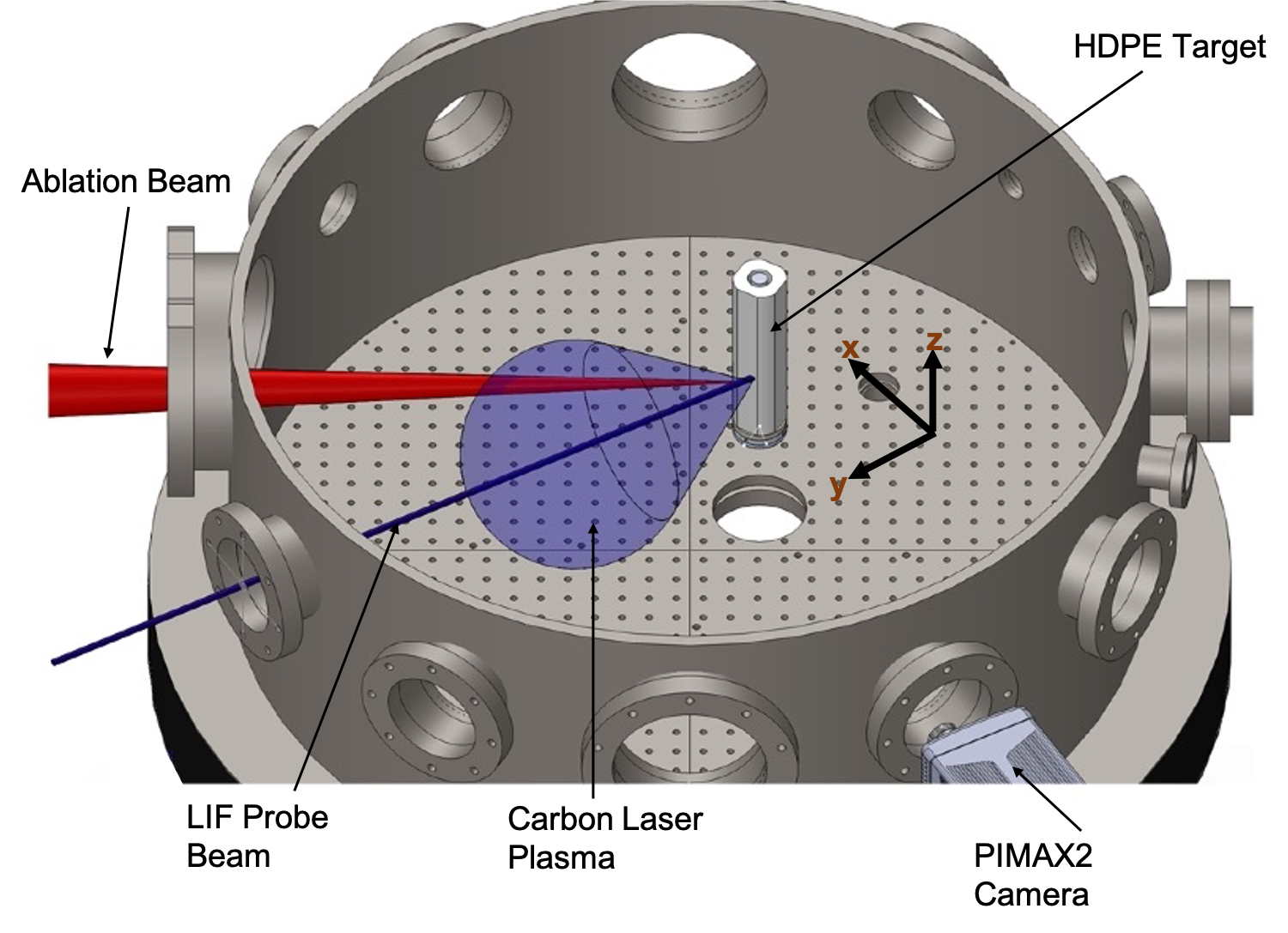}
    \caption{Schematic layout of the experiment. The ablation beam is focused onto the plastic target. The LIF probe beam is directed anti-parallel to the blow-off axis. The PIMAX2 camera images along the -x direction.}
    \label{fig:PI}
\end{figure}

The experiment was performed at the Phoenix Laser Lab \cite{niemann2012} at University of California, Los Angeles (UCLA) using the Peening laser \cite{dane}, a Nd:glass (1.053 ${\mu}m$
wavelength), high-repetition rate (1 Hz), and mid-energy (10 J/pulse) pulsed beam. The 15 ns pulse was focused by a 1 m focal length lens (with a beam f/26) onto the target to a 250 ${\mu}m$ diameter focal spot, at an incidence angle of 34$^{\circ}$ with respect to the y-axis, as shown in Figure~\ref{fig:PI}. 
The target was a 2.54 cm diameter cylindrical rod made of high density polyethylene (HDPE), placed in a $1$ m diameter, $30$ cm tall cylindrical stainless steel chamber, actively pumped down to a 2x10$^{-5}$ Torr vacuum pressure. To ensure a fresh surface at each shot, the target motion was automatically controlled by step motors in helical patterns at 1 Hz, synchronously with the laser and the diagnostics. 

A diode-pumped, solid state, wavelength tunable laser (Ekspla NT230) was used to generate the probe beam for LIF. According to our previous studies \cite{schaeffer_jap}, the predominant charge state in terms of kinetic energy density at these laser intensity levels ($\approx$1$\times$ 10$^{12}$ W/cm$^2$) is C$^{+4}$.  Therefore, the LIF beam was operated around the  $227.091$ nm spectral line of the C$^{+4}$, and at $0.9$ mJ energy per pulse, $4$ ns pulse duration, $50$ Hz. The bandwidth is $6.5$ cm$^{-1}$. 

The LIF beam enters the vacuum chamber through a quartz window, and after intersecting the LPP, terminates onto the target. After the beam intersects the LPP, the volume of ions that are in resonance with the probe beam will fluoresce (cross section $\sim 1$ cm). This determines the spatial resolution of the measurement. The $6.5$ cm$^{-1}$ bandwidth equates to velocity bins of $\approx43$ km/s. However, since the wavelength of the LIF laser can be tuned in $0.01$ nm steps ($\approx13$ km/s at $227.091$ nm) measurements have a $13$ km/s resolution, which is small compared to the $0 - 400$ km/s width of VDF. The intensity of the LIF beam is too low to contribute to the target ablation or heating of the LPP. In this geometry, the LIF diagnostic measured velocities along the y direction by red-shifting from the self emission wavelength in order to resonant with the high velocity ions.

The fluorescing ions are imaged perpendicularly to the probe beam path and blow-off axis, as shown in the experimental setup schematic. The camera-based detection system consists of an image intensified charge-coupled device (ICCD) camera (Princeton Instruments PIMAX2) with a Generation II intensifier sensitive in the ultraviolet (UV) range, an objective ($25$ mm fixed focal length, f/$2.8$, UV sensitive), and a relatively broad ($10$ nm FWHM, $228$ nm central wavelength) optical filter. The optical filter is placed directly in front of the objective to reject stray light from the ablation laser and many spontaneous emission lines. Any spontaneous emission lines that exist within the bandwidth of the optical filters will be subtracted with the background (see Sec.~\ref{sec:img_proc}). 

The fluorescence power saturation was determined by attenuating the LIF beam energy using UV enhanced neutral density filters with optical densities of: $0.1,\ 0.2,\ 0.4$, and $0.7$, corresponding to LIF probe beam energies of $0.7$ mJ, $0.5$ mJ, $0.3$ mJ, and $0.1$ mJ respectively, for a LIF beam wavelength of $227.09$ nm. 

\section{LIF Theory and Analysis}
\label{sec:plif}

\subsection{Feasibility}

One key factor in designing a LIF scheme is having a highly populated bound electron state in which the radiation can resonate with and excite electrons to a higher energy state. Excitation out of the ground state is often preferable since it will often be the most populated state and fluorescence is generally limited by the upper state enhancement. For the carbon ions studied in this experiment, the necessary wavelength to excite an electron out of the ground state is outside the range of commercial lasers ($\lambda_0 \approx 4 - 8$ nm).

Similar to many other He-like ions, there exists a meta-stable state ($1s2s(^2S_1$)) for the electrons to populate in the C$^{+4}$ ion. Excitation out of a metastable state is feasible for LIF, though it will have a much more limited signal gain due to the limited initial population. This is because a captured electron only has a $\approx 5-10\%$ chance of decaying in a way that populates the metastable state\cite{Bonda2012}. This still results in an increased signal-to-noise ratio (SNR) which dominates in comparison to any spontaneous or collisional processes.

\subsection{Two-Level Scheme}

\begin{figure*}
    \centering
    \includegraphics[scale=0.236]{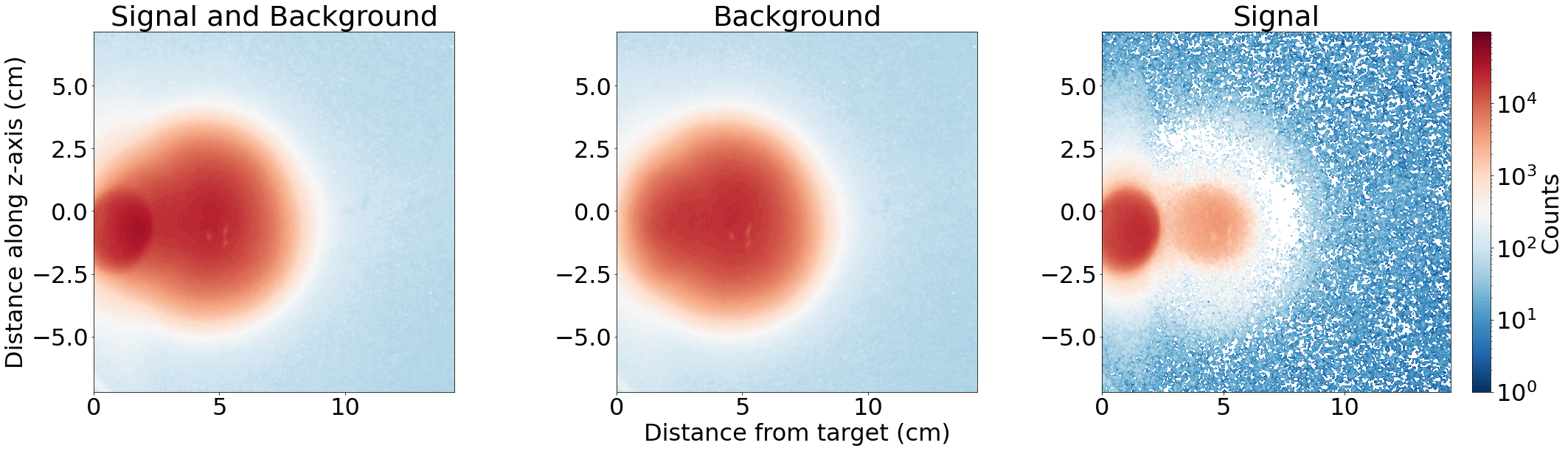}
    \caption{Images taken by the ICCD camera at $100$ ns after the ablation laser fires displaying the background subtraction necessary in the LIF diagnostic: a) the signal and background, averaged over 40 shots; b) the background light only, averaged over 40 shots; c) the difference between the first two panels, which will highlight the fluorescing signal $\approx 5$ cm away from the target surface. The light around $1$ cm is the scattered light from the termination of the LIF beam on the target.}
    \label{fig:SB}
\end{figure*}

Based on the species in question, the population will lend itself to either a two-level or three-level scheme. In the two level scheme the electron transition that we excite is the same one that we will observe. The advantage is that all of the fluorescing ions will decay via the same transition, but it can often present difficulties in terms of scattered light from optics and surrounding metals. This is in opposition to a three level scheme in which the excited electron can then decay via two separate transitions, with the transition to a new state being observed. This offers the advantage of not having to subtract out the scattered light from the LIF probe beam, but without the knowledge of the exact number of electrons in the transition. The  C$^{+4}$ ions we measure undergo a two-level scheme. 

The transition of interest is the $1s2s(^{3}S_{1}) \rightarrow 1s2p(^{3}P_{2})$ for both the absorption and fluorescence of the $C^{+4}$ ions. Excitation to the $1s2p(^{3}P_{0,1})$ states have also been proposed; however, we will only consider the transition to and fluorescence from the $1s2p(^{3}P_{2})$ state due to its  highest statistical weight\cite{Bonda2012}.

LIF on an LPP is challenging due to the the fact that a large portion of the bound electrons are from recombination, and cascade down to the lower bound states. Based on the criterion that the lower state in a LIF scheme must be highly populated compared to the raised state, a lower bound in time is set by how quickly the electrons will recombine and decay to the metastable state. Collisional-radiative modeling suggests that this threshold should be crossed by $\approx 100$ ns after the ablation laser fires\cite{Bonda2012}. The LIF signal will also be brightest where the density is highest. Therefore, an upper bound in time is set by the geometric expansion leading to the dispersion of the LPP density, which for this plasma is $\approx 600$ ns.

\subsection{Saturation and Non-Saturation Regimes}\label{LinSat}

At low probe laser powers, the fluorescence signal is linearly related to the laser power, and therefore the concentration of the absorbing species can be obtained\cite{Altkorn1984}. As laser power is increased, upper level entrancement is eventually limited. This regime is referred to as the saturation limit. Due to experimental conditions, it is often necessary to work at or near the saturation limit. We can further understand this through the rate equations that govern this two-level system

\begin{align}
    \frac{dN_1}{dt} &= -N_1(t)B_{12}I_{\nu}+N_2(t)(A_{21}+Q_{21}+B_{21}I_{\nu})\\
    \frac{dN_2}{dt} &= N_1(t)B_{12}I_{\nu}-N_2(t)(A_{21}+Q_{21}+B_{21}I_{\nu})
\end{align}

\noindent where N$_1$ and N$_2$ represent the population of the two electron levels, B$_{12}$ and B$_{21}$ are the rates of absorption and stimulated emission, A$_{21}$ is the rate of spontaneous emission, Q$_{21}$ is the metastable state quenching rate, and I$_{\nu}$ is the spectral energy density of the LIF probe beam. 

If we assume that I$_{\nu}$ varies slowly in time so that the steady-state condition applies (dN$_{tot}$/dt = 0, where N$_{tot}$ = N$_1$ + N$_2$)  we obtain

\begin{equation}
    N_2 = \frac{B_{12}I_{\nu}}{A_{21}+Q_{21}+B_{21}I_{\nu}}N_1.
\end{equation}

\noindent  In the saturation limit (B$_{12}$I$_{\nu} >>$ c(A$_{21}$ + Q$_{21}$)) the fluorescence is no longer proportional to the laser irradiance and the dominant de-population method is stimulated emission. It has been suggested that in the saturation regime the concentration can be extracted by plotting fluorescence power against inverse laser power, but requires a high intensity probe laser\cite{Bonczyk1979}. This will be further explored in section \ref{sec:satlim}.

\subsection{Image Processing}\label{sec:img_proc}

Light from other sources (background light) must be subtracted so that the intensification of the fluorescing ions can be isolated. In order to account for any long-term experimental changes, background and signal shots were taken in succession in sets of $40-100$ shots, as dictated by the signal-to-noise ratio (SNR), where a shutter blocked the LIF beam every other shot. These were then separated into two groups, signal with background and background, averaged, and subtracted in order to give the average signal. We can see the effects of the background subtraction in Fig. \ref{fig:SB}.

It is important to note that the LIF signal (which only consists of fluorescing C$^{+4}$ ions) has comparatively higher spatial resolution than the background emission, which consists of self-emission from other laser plasma species. The background consists of light from the plane at the lens best focus, as well as defocused light collected along the entire $\approx 10-20$ cm long column of the laser plasma plume (depending on how far the LPP has expanded).

Images produced by the PIMAX 2 cover a spatial area of $23\times23$ cm$^2$ in the plane of best focus. Depending on the data set, the exposure time varied from $2$ ns for the lifetime measurements to $20$ ns for all other shots.

Scattered light (the bright circle seen in Fig.~\ref{fig:SB} (a) and (c)) of the LIF probe beam from the target surface will affect the measurement only in regions within $\approx 2$ cm from the target surface. This prevents certain velocity bins at early time from being accurately measured.

\subsection{Image Sequences}\label{sec:imseq}

\begin{figure}
    \centering
    \includegraphics[scale=0.2]{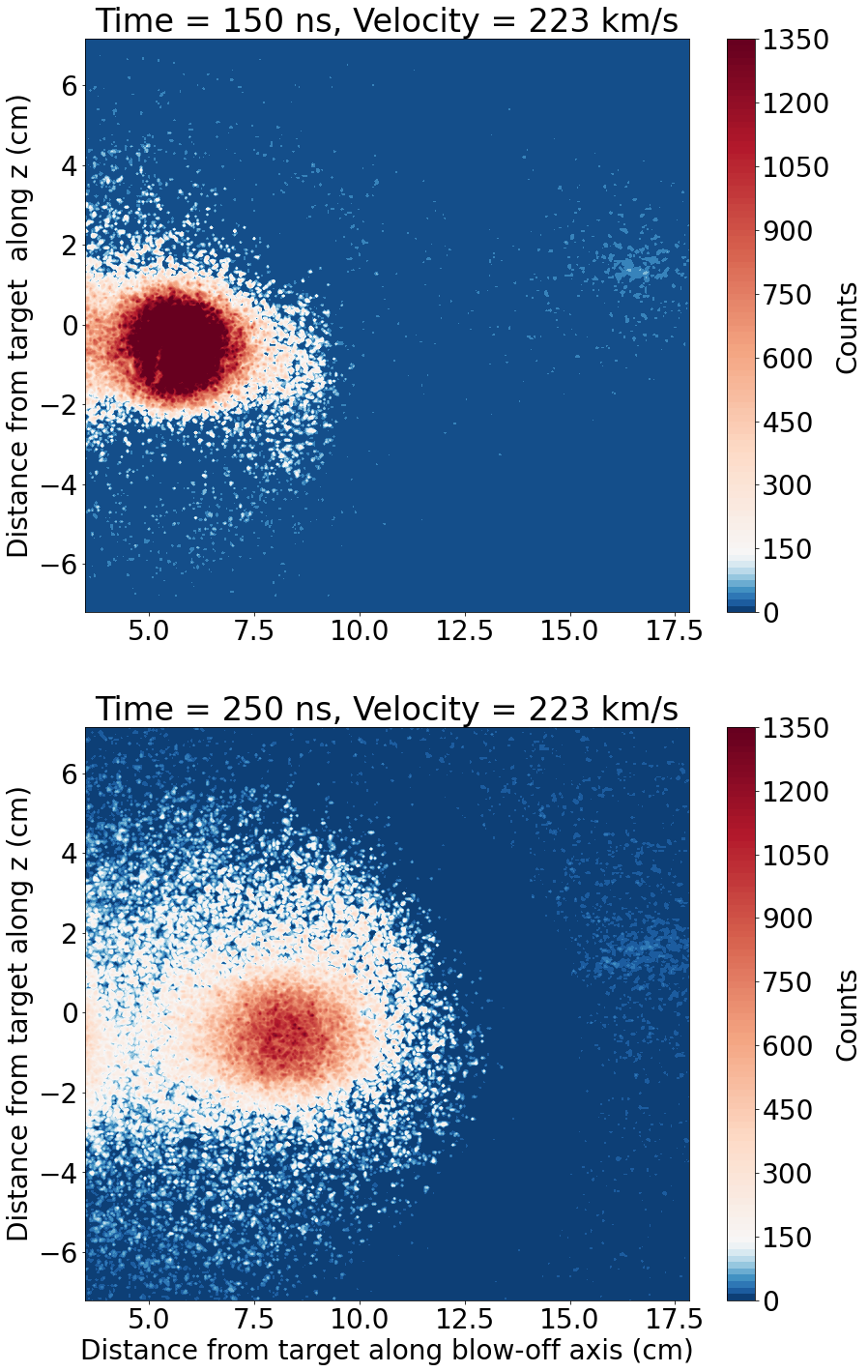}
    \caption{Images of C$+4$ ions traveling at $223$ km/s at two different times relative to the firing of the ablation beam. }
    \label{fig:ImSeq}
\end{figure}

The first image processing stage yields 2D images of fluorescing C$^{+4}$ ions at one set of time and velocity. Example images from the time scans are shown in Fig. \ref{fig:ImSeq}. Panels \ref{fig:ImSeq} (a) and (b) display fluorescing ions moving at $223$ km/s at $150$ and $250$ ns, respectively. These show that the LIF diagnostic is consistent with a ballistic model\cite{dorst2020} for LPP expansion and offers the advantage of not requiring any \textit{a priori} knowledge of the distribution to measure the velocity.

Each sequence of shots (i.e. wavelength scan at a constant time or time scan at a constant wavelength) can be combined into a streak plot in order to represent the evolving ion dynamics in the system. For this purpose, we average the signal across the z-axis to reduce each image to a $1$D array and stack the arrays along the scanned parameter. For the velocity scans we reduce the data down to the phase space velocity plot (Fig. \ref{fig:TS_VDF} (b)) and for the time scans we get streak plots of the spatio-temporal evolution of a single velocity bin (Fig. \ref{fig:TS_VDF} (a)). 

\section{Results}
\label{sec:results}

\subsection{Spatio-Temporal Evolution Maps and Velocity Distribution Function}\label{sec:maps}

\begin{figure}
    \centering
    \includegraphics[scale=0.55]{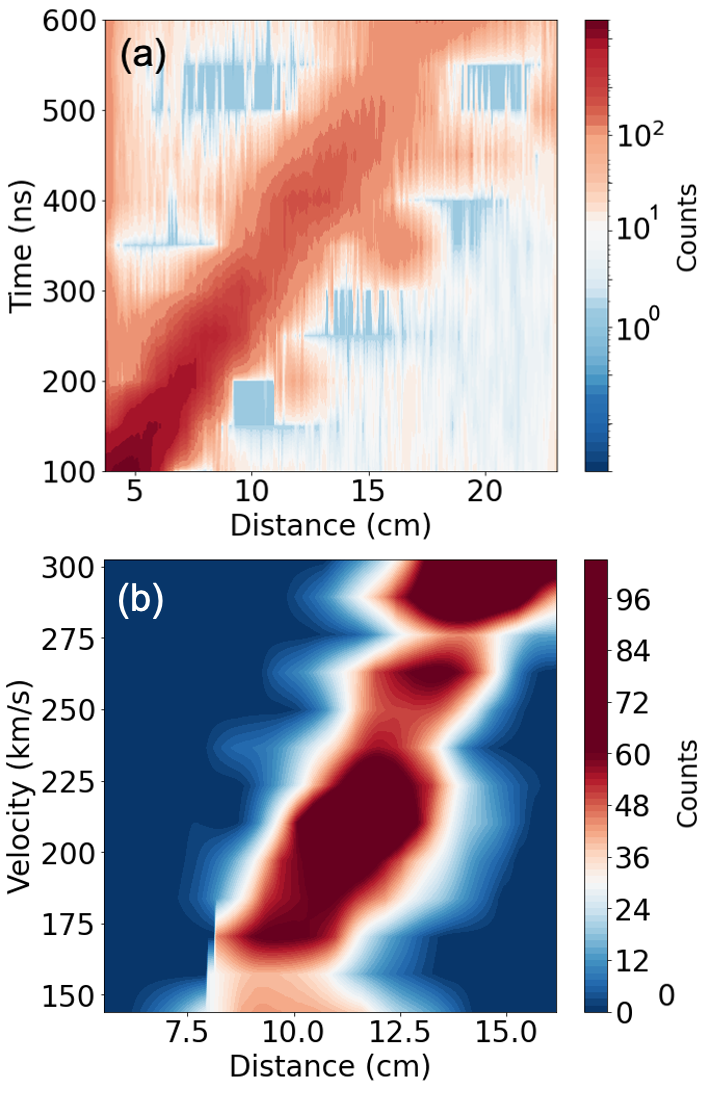}
    \caption{(a) Streak map of the spatio-temporal evolution of the C$^{+4}$ ions.  The illuminated velocity bin is around $300$ km/s. (b) A VDF map taken at t = 400 ns relative to the target irradiation. The data is consistent with a time-of-flight model for a freely-expanding LPP.}
    \label{fig:TS_VDF}
\end{figure}

The spatio-temporal evolution map in Fig. \ref{fig:TS_VDF} (a) shows the evolution of C$^{+4}$ traveling between $280$ - $320$ km/s (LIF beam tuned to $227.32$ nm). In this experiment, since there is no external magnetic fields, plasma, or gas to interact with, we observe the slope of the spatio-temporal evolution map to remain constant and correspond to $\approx300$ km/s as expected. 

An example of a measured VDF is displayed in Fig. \ref{fig:TS_VDF} (b). The data was acquired at t = $400$ ns after the ablation beam fired, and the LIF beam was scanned through $150 - 300$ km/s. At velocities lower than $150$ km/s, the signal was obscured by the scattered light. The signal extended beyond the field-of-view for velocities greater than $300$ km/s. 

\subsection{Saturation Limit}\label{sec:satlim}

\begin{figure}
    \centering
    \includegraphics[scale=0.3]{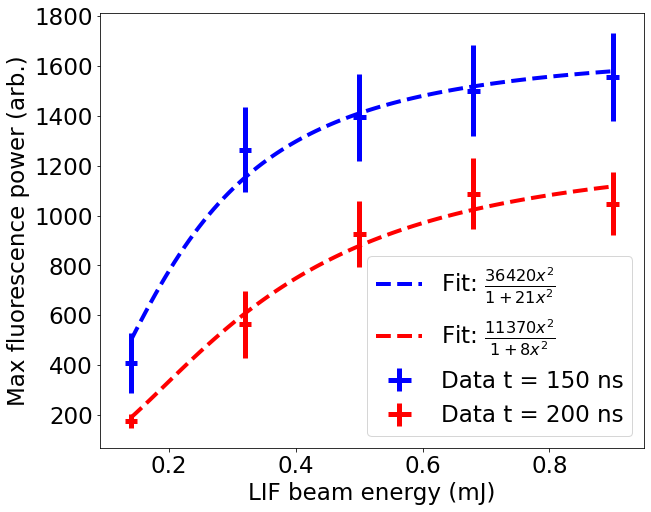}
    \caption{Scaling of the C$^{+4}$ fluorescence with the LIF laser intensity, 
    shown at 150 ns (blue) and 200 ns (red). }
    \label{fig:IS}
\end{figure}

In order to assess whether the diagnostic operates in the linear or saturation regime, the LIF probe beam energy was scanned through five energies: $0.1,\ 0.3,\ 0.5,\ 0.7,\and 0.9$ mJ. Figure \ref{fig:IS} shows the effects of the intensity scan on fluorescence power (intensity) of the C$^{+4}$ population at two different times. 

At $150$ ns (the blue data and fit) the fluorescent power saturates above $0.5$ mJ laser energy, as varying the intensity of the LIF beam does not affect the fluorescence. At lower energies the system exhibits non-saturation effects. In order to extract the concentration of the species we would have to conduct a more thorough intensity scan, or have a higher energy probe beam.

By $200$ ns (the red data and fit) the laser plasma has freely expanded and fewer points in the intensity scan are saturated. There is a noticeable linear region (at least three points), which is ideal when extracting concentration. These two preliminary scans show that concentration measurements are viable with this diagnostic. 

\subsection{Lifetime Measurement}

\begin{figure}
    \centering
    \includegraphics[scale=0.3]{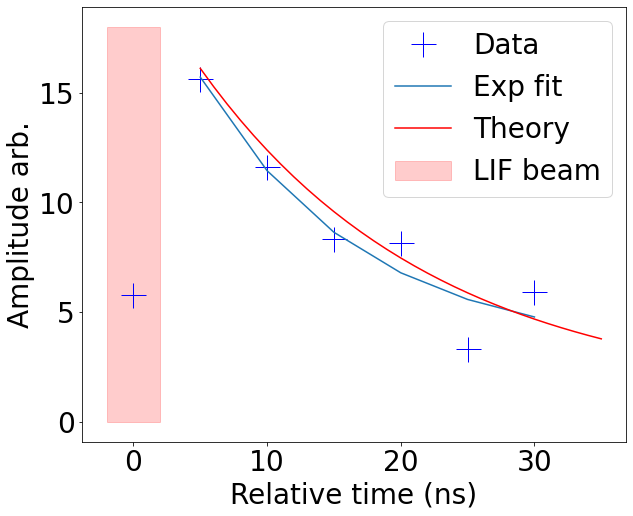}
    \caption{The lifetime of the $1s2p(^3P_2)$ state was measured. After the LIF beam resonates electrons from the metastable $1s2s(^2S_1)$ to the upper states, the decay rate of measured in successive shots and fit with an exponential decay curve. The theoretical lifetime curve is plotted for comparison. }
    \label{fig:Life}
\end{figure}

The fluorescence lifetime is the amount of time the ion will remain in an excited state before decaying to the lower state. After optical pumping, this decay rate can be measured by finely changing the delay between the LIF probe beam and the camera exposure. This requires changing the exposure time of the camera to $2$ ns, and scanning in steps of $2$ ns. 

We can see the results of this measurement in Fig.~\ref{fig:Life}. The theoretical lifetime (plotted in red) has a 1/e lifetime of 17.64 ns\cite{Cann1992}. The fitted lifetime (blue) was measured to be $\approx 19$ ns. 

\section{Conclusion}\label{sec:summary}

We have developed a new use-case of a planar laser induced fluorescence diagnostic for mapping the spatio-temporal evolution of the velocity distribution function in an explosive carbon laser produced plasma. The relatively large bandwidth of the LIF probe beam (6.5 cm$^{-1}$) suited the measurement of the wide velocity distribution of the laser plasma ($\approx 42\ cm^{-1}$) in a realistic number of shots for a high repetition rate facility. Each $1$D profile was the collection of $40 - 100$ laser shots, making each dataset upward of $1000$'s of shots. This sort of measurement would not be feasible in single shot experiments. We were also able to asses the saturation limit of the the laser plasma with regards to the LIF probe beam irradiance. The lifetime of the C$^{+4}\ 1s2p(^3P_2)$ state was also verified and found to be consistent with previous measurements. 

\section{Acknowledgements}

This work was supported by the Defense Threat Reduction Agency, Lawrence Livermore National Security LLC under Contract No. B649519, the United States Department of Energy (DOE) under Contract No. DE-SC0021133, and and the National Science Foundation Graduate Fellowship Research Program (Award No. DGE-1650604). We thank NIWC Pacific and Curtiss-Wright MIC for help with the slab laser.

\section{Data Availability Statement}

The data that support the findings of this study are available from the corresponding author upon reasonable request.
 
\section{References} 
%
\bibliography{LIF.bib}

\end{document}